\documentclass[a4paper,11pt]{article}
\usepackage{pos}

\title{Aspects of massive gauge fields}

\author*[a]{Anamaria Hell}

\affiliation[a]{Kavli IPMU (WPI), UTIAS, The University of Tokyo,\\ Kashiwa, Chiba 277-8583, Japan}

\emailAdd{anamaria.hell@ipmu.jp}

\abstract{Massive gauge fields whose mass is introduced by hand form very intriguing theories. They depart from their massless counterparts by a straightforward modification. Yet, taking the limit when the same vanishes poses a non-trivial challenge. In these notes, with a focus on vector and two-form fields, we will discuss several aspects that arise when one explores the massless limit. We will study new connections among different theories, and at times raise a question about the already established ones.   }

\FullConference{Proceedings of the Corfu Summer Institute 2024 "School and Workshops on Elementary Particle Physics and Gravity" (CORFU2024)\
12 - 26 May, and 25 August - 27 September, 2024\\
Corfu, Greece\\}

\begin{document}
\maketitle

\section{Introduction}

Adding a mass term by hand is among the simplest modifications that one could make when studying vector fields, gravity, or more general theories. It is also tempting to do so, as observational data provides an upper bound on the mass of the particles, such as, for example, the photon. Moreover, it is also increasingly appealing in the light of dark energy and dark matter -- unknown components that give rise to the late-time acceleration of the universe, and account for its missing mass. 
Nevertheless, sometimes even the simplest modifications can lead to the most surprising conclusions. 
According to the principle of continuity, introduced in \cite{BassSch}, all of the observables of the modified theory should match with those of the original theory when the mass is set to zero. However, historically, this didn't necessarily seem to be always the case.  
\\\\
Massive Yang-Mills theory (mYM), a theory of non-Abelian vector fields whose mass is added \textit{"by hand"} to the Lagrangian was originally introduced to explain the weak interactions, serving as one of the starting points in the construction of the Standard model of particle physics \cite{Glashow}. It is also interesting from the cosmological point of view as an inflationary candidate and was suggested to resolve the hierarchy problem \cite{Delburgo, Nieto:2016gnp}. It has a seemingly simple modification when compared to its massless counterpart -- the action between the two differs only by mass. However, in contrast to the massless theory, or even the theory in which the mass is introduced through the Higgs mechanism, which is perturbatively renormalizable and unitary \cite{masslessYM, HiggsYM},  this theory appears to be significantly more complicated. In particular, its propagator tends to a constant at high energies: 
\begin{equation}\label{eq::propagator}
    \tilde{\Delta}_{\mu\nu}^{ab}(k)=\left(-\eta_{\mu\nu}+\frac{k_{\mu}k_{\nu}}{m^2}\right)\frac{i\delta^{ab}}{k^2-m^2+i\epsilon}.
\end{equation}
and thus indicates a power-counting non-renormalizable theory. Moreover, in contrast to the other two theories, the perturbative series in (mYM) appear to be singular in mass, thus suggesting the violation of unitarity at energy-scales \cite{ SalamKomar, UmezawaKamefuchi, VeltmanReiff}: 
\begin{equation}
    k_g\sim\frac{m}{g},
\end{equation}
Where $g$ is the coupling constant. Notably, in \cite{Veltman1970}, it was noticed that the singularities in mass form a series of the form $\frac{g^2\Lambda^2}{m^2}$, where $\Lambda$ is the cut-off scale, raising the possibility that the series might be re-summed such that the singularities may be cured. However, the same year this possibility was dismissed in \cite{vDVZ} -- by studying the one-loop contributions to the propagator of the transverse modes, it was noticed that even though this value is finite, the imaginary part of the diagrams differs from the massless case by a factor of $1/2$, which remains in the massless limit. As argued in \cite{vDVZ}, this is due to the longitudinal modes, extra degrees of freedom that appear in the mYM theory in and are absent in the massless version, which fail to decouple from the remaining transverse modes. 
\\\\
MYM theory was not the only theory that faced this issue -- as shown in \cite{Zakharov, vDVZ}, if one takes linearized gravity, and adds to it a mass term of the Fierz-Pauli form \cite{FierszPauli}, the predictions of such theory for the deflection of starlight or precession of perihelia of Mercury would differ from those of standard linearized gravity to such an extent that one could disregard the massive theory as a possible theory of nature. The reason is the same -- the mass term introduces an additional degree of freedom. Among them, the longitudinal mode, which is absent in the massless theory fails to decouple in the massless limit, thus giving rise to the apparent discontinuity. Yet, as pointed out in \cite{VainshteinMeh}, massive gravity is a non-linear theory, and thus, there should be a scale, called \textit{the Vainstein radius}, at which the perturbative expansion for this mode breaks down due to the non-linear terms. As a result of the Vainshtein mechanism, the apparent discontinuity is just an artifact of the perturbative approach -- beyond the radius, the strongly coupled longitudinal mode decouples from the remaining modes thus recovering general relativity up to small corrections \cite{PertNep, Gruzinov}.\footnote{It should be noted that in addition, the mass can give rise to the additional 6th degree of freedom that could be a ghost \cite{Boulware:1972yco}. However, there have been constructions that claim to avoid this problem (see eg. \cite{ deRham:2010kj, Mimetic}). }Moreover, as shown in the case of mimetic massive gravity, a similar effect appears for the vector modes as well \cite{Mimetic} which is also absent in the massless theory, in addition to the longitudinal mode.  
\\\\
This brings an exciting possibility that the apparent discontinuity for the mYM theory might be resolved in a similar way, outside of the perturbation theory, as conjectured already in \cite{Vainshtein}. In these notes, based on the works in \cite{Hell:2021wzm, Hell:2021oea, Hell:2022wci, Hell:2024xbv}, and an invited plenary talk at the CORFU2024, The Dark Side of the Universe (DSU2024),  we will summarize the results that show that this is indeed the case, and further discuss the consequences of exploring the massless limit to Abelian vector and two-form fields with self-interactions and non-minimal coupling, discussing the aspects of such theories. 
In particular, the notes are structured as follows: First, in section 2 we will discuss the massless limit of mYM theory. Then, in section 3, we will focus on two other theories, of massive vector and 2-form fields, comparing the results with the previous case and discussing the possible connections among them in the presence of self-interactions. Then, in section 3, we will focus on Proca theory in the presence of non-minimal coupling to gravity, and find curious inconsistencies as well as connections to the mYM theory. Finally, we will conclude with a short summary of the notes. 

\section{The massless limit of massive Yang-Mills theory}\label{secmYM}
As a first step in these notes, we will analyze the massive Yang-Mills theory (mYM), outlining the main results for its massless limit, based on the works \cite{Hell:2021oea, Hell:2022wci}. This is a theory of non-Abelian vector fields $A_{\mu}$, given by the $2\times2$ hermitian matrices for the $SU(2)$ theory,  described by the following action: 
\begin{equation}\label{mYMaction1}
    S=\int d^4x \left[-\frac{1}{2}\text{Tr}(F_{\mu\nu}F^{\mu\nu})+m^2\text{Tr}(A_{\mu}A^{\mu})\right].
\end{equation}
Here,  
\begin{equation}
      F_{\mu\nu}=D_{\mu}A_{\nu}-D_{\nu}A_{\mu},\qquad\text{and}\qquad   D_{\mu}=\partial_{\mu}+igA_{\mu},
\end{equation}
is the field strength tensor and covariant derivative respectively. One can decompose the vector field in terms of the generators of the SU(2) group:  
\begin{equation}
    A_{\mu}=A_{\mu}^aT^a,    \qquad \text{with}\qquad T^a=\frac{\sigma^a}{2} \qquad\text{and}\qquad a=1,2,3.
\end{equation}
 To gain insight into the behavior of this theory, our goal will be first to rewrite it only in terms of its propagating modes. The vector field has four components -- the temporal component and three spatial ones. However, by separating the vector field in terms of them, one can notice that in contrast to the spatial part, the temporal does not show up in a standard form, lacking a term such as $\dot{A}_0\dot{A}_0$ with dot denoting a derivative with respect to time. Rather, this component is constrained. So to find only the dof explicitly, one should find this constraint, solve it, and substitute it back to the action. Before we continue with this procedure, however, two comments are in order. First, one should note that commonly, massive gauge theories are rewritten \textit{a la Stueckelberg}, especially in the context of particle physics. This introduces an additional gauge redundancy in the system, and one should then fix the gauge per the standard description. Here, we will follow an approach that is more similar to the cosmological perturbation theory, decomposing instead the field in terms of the scalar and vector modes, even though we we are working in a flat space-time background. In this way, we avoid any potential fictitious modes through Lorentz-covariant gauge\footnote{See \cite{Hell:2023mph} on how the Lorentz-like gauge can give rise to fictitious modes in certain vector or gravity theories.}, and at the same time write the theory only in terms of the components that are propagating in the theory having thus a direct overview of the entire behavior of the modes. Secondly, it is best to decompose the spatial part of the vector field into the transverse and longitudinal modes in the following way: 
\begin{equation}\label{eq::Nonlindec}
    A_i=\zeta A_i^T\zeta^{\dagger}+\frac{i}{g}\zeta_{,i}\zeta^{\dagger},\qquad \text{with}\qquad A_{i,i}^T=0.
\end{equation}
Here, the comma denotes a derivative with respect to the spatial component. We will refer to the $A_i^T$ as the transverse modes. The unitary matrix $\zeta$ carries the longitudinal modes, defined in the following way: 
\begin{equation}
    \zeta=e^{-ig\chi}. 
\end{equation}
By expanding the above expression, and writing in terms of the generators of $SU(2)$, we find: 
\begin{equation}\label{nlindecEXP}
    \begin{split}
        A_i^a=A_i^{Ta}+\chi_{,i}^a-g\varepsilon^{abc}(A_i^{Tb}\chi^c+\frac{1}{2}\chi_{,i}^b\chi^c)-\frac{g^2}{2}\varepsilon^{fab}\varepsilon^{fcd}(\chi^bA_i^{Tc}\chi^d+\frac{1}{3}\chi^b\chi^c_{,i}\chi^d).
    \end{split}
\end{equation}
Usually, when decomposing the modes into scalar and vector ones, one would keep only the first two linear terms in the above decomposition. Here, on the other hand, we have also introduced the non-linear terms. The reason for this is two-fold. First, as shown in \cite{Hell:2021oea}, it is possible to analyze the theory just by considering the first two terms. However, at the end of the computation, one will necessarily come to the non-linear decomposition. We will further elaborate on this detail at a later point. Secondly, the transverse modes defined in (\ref{nlindecEXP}) are gauge invariant to all orders in perturbation theory, in the corresponding massless case. Thus, as shown in \cite{Hell:2022wci}, the above decomposition allows us to write the theory only in terms of the transverse modes after all constraints are resolved, which would not be possible for the linear decomposition -- the longitudinal modes would remain at the third order action in perturbation theory. 
\\\\
By varying the action with respect to the temporal component, and decomposing the spatial one according to (\ref{nlindecEXP}), we find the following constraint: 
\begin{equation}\label{eq::constraint1}
    \begin{split}
        \left(-\Delta+m^2\right)A_0&=-\dot{A}_{i,i}+ig[\dot{A}_i,A_i]
        +ig\left(2\left[A_i,A_{0,i}\right]+\left[A_{i,i},A_0\right]\right)+g^2\left[A_i,\left[A_0,A_i\right]\right],
    \end{split}
\end{equation}
whose solution is given by: 
\begin{equation}\label{eq::Csolution}
    A_0=\zeta\frac{1}{D}\left(-\frac{i}{g}m^2\zeta^{\dagger}\dot{\zeta}+ig\left[\dot{A}_i^T,A_i^T\right]\right)\zeta^{\dagger}+\frac{i}{g}\dot{\zeta}\zeta^{\dagger}.
\end{equation}
Here,
\begin{equation}
    \frac{1}{D}=\frac{1}{-\Delta+m^2-2ig[A_i^T,\partial_i\bullet\;]+g^2[A_i^T,[A_i^T,\bullet\;]]}
\end{equation}
is an operator, with $\bullet$ denoting a place where expression on which $\frac{1}{D}$ acts should be placed, when perturbatively evaluated. Substituting this back to the action, and expanding, we find the following leading-order contributions for energies $k^2\gg m^2$: 
\begin{equation}\label{eq::Ldenspert}
    \mathcal{L}=\mathcal{L}_{0}+\mathcal{L}_{int}, \qquad\text{where}
\end{equation}
\begin{equation*}
    \begin{split}
        \mathcal{L}_0&=\text{Tr}\left[-\chi\left(\Box+m^2\right)\frac{-\Delta m^2}{-\Delta+m^2}\chi-A_i^T\left(\Box+m^2\right)A_i^T\right]\\
        \mathcal{L}_{int}&\sim\text{Tr}\left\{igm^4[\chi,\dot{\chi}]\frac{1}{\Delta}(\dot{\chi})-2igm^2A_i^T\chi\chi_{,i}-2ig\left[\dot{A}_i^T,A_i^T\right]\frac{m^2}{\Delta}\left(\dot{\chi}\right)-2igA_i^TA_j^T(A_{j,i}^T-A_{i,j}^T)\right.\\
        &\left.+\frac{m^2g^2}{6}\left(\chi_{,\mu}\chi\chi^{,\mu}\chi-\chi_{,\mu}\chi^{,\mu}\chi^2\right)-\frac{g^2m^2}{3}A_i^T\left[\chi,\left[\chi_{,i},\chi\right]\right]\right\}.
    \end{split}
\end{equation*}
The free part of the Lagrangian shows that this theory describes three degrees of freedom (dof) -- a longitudinal and two transverse modes. Moreover, we can see that by setting $m^2=0$, the resulting theory matches with the massless case -- the longitudinal mode disappears from the action. At the same time, however, we can notice that the longitudinal modes are not canonically normalized. By taking into account that the minimal amplitude of quantum fluctuations for the normalized field $\phi$ is given by $\delta\phi^2\sim\frac{k^3}{\omega_k}$ for length scales $L\sim\frac{1}{k}$, where $\omega_k$ is determined by the dispersion relation, we infer that the longitudinal and transverse modes for high energies $k^2\sim\frac{1}{L^2}\gg m^2$ can be evaluated as: 
\begin{equation}
     \delta\chi_L\sim\frac{1}{mL}\qquad\text{and}\qquad  \delta A_L^{T}\sim\frac{1}{L}. 
\end{equation}
Therefore, while at first order in the perturbation theory, the terms are finite, at $\mathcal{O}(g^2)$, the longitudinal modes give rise to singular-in-mass behavior: 
\begin{equation}
   \frac{m^2g^2}{6}\left(\chi_{,\mu}\chi\chi^{,\mu}\chi-\chi_{,\mu}\chi^{,\mu}\chi^2\right)\sim \frac{g^2m^2\chi^4}{L^2}\sim\frac{g^2}{\left(mL\right)^2L^4}
\end{equation}
Here, we have estimated the derivatives with $\partial_{\mu}\sim\frac{1}{L}$. In particular, one can check that the above term will be the most dominant, among all possible non-linear terms. On the level of the equation of motion, it contributes to the second-order corrections: \begin{equation}
    \left(\Box+m^2\right)\chi^{(2)}\sim\frac{g^2}{3}\left[\chi^{(0)}_{,\mu},\left[\chi^{(0)},\chi^{(0),\mu}\right]\right]\sim\frac{g^2}{\left(mL\right)^3L^2}\sim\frac{g^2}{\left(mL\right)^3}.
\end{equation}
Here, we have perturbatively expanded the longitudinal and transverse modes as: 
\begin{equation}
    \chi=\chi^{(0)}+\chi^{(1)}+...\qquad\text{and}\qquad A_i^T=A_i^{T(0)}+A_i^{T(1)}+...
\end{equation}
such that the $\chi^{(0)}$ and $A_i^{T(0)}$ satisfy the free equations. Once the second-order corrections become of the same order as the linear term $\chi^{(0)}\sim\frac{1}{mL}$, the perturbation theory in the longitudinal mode breaks down. The scale which characterizes this breakdown is called the strong coupling scale: 
\begin{equation}
    L_{str}\sim\frac{g}{m},
\end{equation}
and it matches exactly with the unitary scale. Therefore, for length- scales larger than the strong coupling scale, one can apply the perturbative approach for the longitudinal modes. However, the same stops to hold once $\mathcal{L}_{str}$ is reached. One can verify that at this point, the constraint for the temporal component can still be resolved. However, one can no longer expand the matrix $\zeta$. Thus, the Lagrangian (\ref{eq::Ldenspert}) is no longer a good approximation, and should be replaced with: 
\begin{equation}\label{eq::nonpertLagrangian}
    \mathcal{L}=\mathcal{L}^T_0+\mathcal{L}_0^{\chi}+\mathcal{L}^T_{int}+\mathcal{L}^{T\chi}_{int}
\end{equation}
\begin{equation*}
    \begin{split}
        &\mathcal{L}^T_0=\text{Tr}\left(\dot{A}_i^T\dot{A}_i^T-A_{i,j}^TA_{i,j}^T-m^2A_i^TA_i^T\right)\\\\
        &\mathcal{L}_0^{\chi}=-\frac{m^2}{g^2}\text{Tr}\left[\zeta^{\dagger}\dot{\zeta}\frac{-\Delta}{-\Delta+m^2}\left(\zeta^{\dagger}\dot{\zeta}\right)-\zeta^{\dagger}\zeta_{,i}\zeta^{\dagger}\zeta_{,i}\right]\\\\
        & \mathcal{L}^{T\chi}_{int}=\frac{2im^2}{g}\text{Tr}\left\{-A_i^T\zeta^{\dagger}\zeta_{,i}+m^2\zeta^{\dagger}\dot{\zeta}\frac{1}{D}\left[A_i^T,\frac{1}{-\Delta+m^2}\partial_i(\zeta^{\dagger}\dot{\zeta})\right]\right\}\\\\
        &-m^2\text{Tr}\left\{\zeta^{\dagger}\dot{\zeta}\frac{1}{D}[\dot{A}_i^T,A_i^T]+[\dot{A}_i^T,A_i^T]\frac{1}{D}(\zeta^{\dagger}\dot{\zeta})+m^2\zeta^{\dagger}\dot{\zeta}\frac{1}{D}\left[A_i^T,\left[A_i^T,\frac{1}{-\Delta+m^2}(\zeta^{\dagger}\dot{\zeta})\right]\right]\right\}\\\\
        &\mathcal{L}^T_{int}=\text{Tr}\left\{-2igA_i^TA_j^T\left(A_{j,i}^T-A_{i,j}^T\right)+g^2\left[\dot{A}_i^T,A_i^T\right]\frac{1}{D}\left[\dot{A}_j^T,A_j^T\right]+g^2\left(A_i^TA_j^TA_i^TA_j^T-A_i^TA_i^TA_j^TA_j^T\right)\right\}.
    \end{split}
\end{equation*}
Therefore, for scales $L\leq L_{str}$, the new kinetic term for the longitudinal modes is given by $\mathcal{L}_0^{\chi}$. By performing several field redefinitions, and canonically normalizing the resulting fields as in \cite{Hell:2021oea}, taking into account that  $\mathcal{L}^T_0$ is still the dominant kinetic term for the transverse modes, indicating that they can be evaluated as $A_L^{T}\sim\frac{1}{L}$,  it is possible to show that the leading order corrections to the transverse modes due to the longitudinal ones are given by: 
 \begin{equation}
    A_i^{T(1)}\sim-i\frac{m^2}{g}\zeta^{\dagger}\zeta_{,i}\sim\frac{g}{L^3}\frac{L}{L_{str}}.
\end{equation}
Let us now take the massless limit. As mass goes to zero, the strong coupling scale goes to infinity. Thus, the above ratio becomes smaller as we approach higher energies, ultimately vanishing in the massless limit. This shows thus that similarly to the Vainshtein mechanism in massive gravity, the apparent discontinuity in mass is just an artifact of the perturbation theory.  The massless limit in massive Yang-Mills theory is smooth, with longitudinal modes completely decoupling from the remaining degrees of freedom, which remain weakly coupled for all scales. 
\\\\
Curiously, by comparing mYM theory and massive gravity, we can notice a trend -- degrees of freedom that are absent in the massless theories, become strongly coupled at a certain length scale in the massive case, when self-interactions are taken into account. This can even be extended to $R^2$ gravity if one considers the theory in flat space-time \cite{Hell:2023mph}.  As we will see in the next section, this has indications for relations among massive gauge theories, known as dualities. 

\section{Proca \textit{vs.} the Kalb-Ramond fields}\label{dualities}
In the previous section, we have considered mYM theory, which described non-Abelian fields. In this section, we will instead consider a simpler Abelian case, focusing on vector and two-form fields. In particular, the first theory that we will be interested in is the Proca theory -- the theory of a massive vector field \cite{Proca}.  Similarly to the non-Abelian case, this theory describes three dof -- a longitudinal mode, and two transverse ones. The transverse modes are also present in its massless counterpart, known as electrodynamics. The Kalb-Ramond (KR) theory, in contrast, is a theory of an antisymmetric two-form \cite{KalbRamond}. In the massless case, it describes a single, pseudoscalar dof. However, curiously, if taken massive, it describes three dof as well -- a pseudo scalar longitudinal dof and two transverse modes. While the two theories essentially deal with different fields, and allow for different couplings to the external matter due to the scalar-pseudo-scalar nature, there have been numerous claims in the literature that the two are dual, meaning that they describe the same physics (see eg.\cite{Kawai, Trugenberger, Quevedo, 2001Smailagic, 2002Casini, Auria, Buchbinder, Dalmazi, Shifman, Garcia, Kuzenko, 2002string, Zinoviev, McReynolds, 2017Aurilia, 2019Markou, 2019Schmidt, Heisenberg, 2020SmajlSpal, Barbosa:2022zfm}). This is however curious from the point of view of the principle of continuity -- if massless, one theory describes a single dof, whereas the other two. For the infinitely small mass, the massless and massive theories should thus match, which does not appear to describe the same physical effects. Therefore, following the analog of the Vainshtein mechanism for the previous theories, one could expect that in Proca theory, the perturbation theory would break for the longitudinal modes, as already shown to be the case for some interactions in \cite{Khoury, GenPV}, whereas, for the Kalb-Ramond field, this would be the case for the two transverse modes. 
\\\\
To explore this further, following \cite{Hell:2021wzm}, in this section, we will analyze the Kalb-Ramond and Proca theories in the presence of quartic self-interactions. In particular, we will consider: 
\begin{equation}\label{eq::KBactionInt}
    S=\int d^4x\left[\frac{1}{12}H_{\mu\nu\rho}H^{\mu\nu\rho}+\frac{m^2}{4}B_{\mu\nu}B^{\mu\nu}+\frac{g^2}{16}\left(B_{\mu\nu}B^{\mu\nu}\right)^2\right],
\end{equation}
and
\begin{equation}\label{eq::PactionInt}
    S=\int d^4x\left[-\frac{1}{4}F_{\mu\nu}F^{\mu\nu}+\frac{m^2}{2}A_{\mu}A^{\mu}+\frac{g^2}{4}\left(A_{\mu}A^{\mu}\right)^2\right].
\end{equation}
The first action describes the Kalb-Ramond field $B_{\mu\nu}$, which is antisymmetric in the two indices, while the second action corresponds to the Proca theory, where the vector field $A_{\mu}$ is now Abelian. The 
\begin{equation}
     H_{\mu\nu\rho}=B_{\nu\rho,\mu}+B_{\rho\mu,\nu}+B_{\mu\nu,\rho}\qquad \text{and}\qquad F_{\mu\nu}=\partial_{\mu}A_{\nu}-\partial_{\nu}A_{\mu}
\end{equation}
are the corresponding field strengths. As a next step, let's consider the behavior of the fields in the two theories. For this, we will decompose the fields according to the irreducible representations of the $SO(3)$ group. In particular, for the KR field, we have \cite{HermGrav}:
\begin{equation}
    \begin{split}
        &B_{0i}=C_i^T+\mu_{,i},\qquad\text{}\qquad C_{i,i}^T=0\\\\
        &B_{ij}=\varepsilon_{ijk}B_k,\qquad\qquad B_i=B_i^T+\phi_{,i},\qquad\text{}\qquad B_{i,i}^T=0
    \end{split}
\end{equation}
whereas the Proca field can be decomposed as: 
\begin{equation}\label{lindecP}
    \begin{split}
        &A_0\\
        &A_i=A_i^T+\chi_{,i},\qquad\qquad A_{i,i}^T=0. 
    \end{split}
\end{equation}
Similarly to the mYM theory, Proca theory has only one constraint, which can be found by varying the action with respect to the temporal component. The KR field, on the other hand, has two, which can be found by varying the action with respect to two $C_i^T$ and $\mu$. Once we solve these constraints and substitute back to the actions, we find for $k^2\sim\frac{1}{L^2}\gg m^2$:
\begin{equation}
    \begin{split}
        \mathcal{L}_P\sim-\frac{1}{2}\chi(\Box+m^2)\frac{-\Delta m^2}{-\Delta+m^2}\chi-\frac{1}{2}A_i^T(\Box+m^2)A_i^T+\frac{g^2}{2}(\chi_{,\mu}\chi^{,\mu})^2-g^2\chi_{,\mu}\chi^{,mu}\chi_{,i}A_i^T
    \end{split}
\end{equation}
for the Proca field, and 
\begin{equation}\label{KRpert}
    \mathcal{L}_{KR}\sim-\frac{1}{2}B_i^T(\Box+m^2)\frac{ m^2}{-\Delta+m^2}B_i^T-\frac{1}{2}\phi(\Box+m^2)(-\Delta)\phi+g^2(B^T)^4+g^2(B^T)^3\frac{\phi}{L}
\end{equation}
for the Kalb-Ramond one. In the above expressions, we have only considered the most important self-interactions, among infinitely many. In addition, for simplicity, we have schematically written those for the KR field, estimating $\partial_{\mu}\sim\frac{1}{L}$. We can see that the crucial differences between the two lie in the kinetic terms. On the one hand, in the Proca case, the longitudinal mode is multiplied by the mass term. Thus, if canonically normalized, and perturbative, it will cause the appearance of terms that are singular in mass. This is similar to the previous case in the mYM theory -- the longitudinal mode was absent in the massless theory and thus appeared multiplied by mass in the massive case. In contrast, the massless KR theory has only a pseudo-scalar mode, that corresponds to $\phi$. The transverse modes are absent, and thus appear multiplied by the mass in the massive theory. However, then, once one canonically normalized the theory, it is clear that these two modes will give rise to the terms that are singular in mass. 
Following \cite{Hell:2021wzm}, by studying the two theories on the level of the equation of motion, and taking into account their minimal amplitude of quantum fluctuations, 
we can find a similar effect as in the mYM case. In the Proca theory, the longitudinal mode becomes strongly coupled at the scale $ L_{str}\sim\frac{\sqrt{g}}{m}. $ Beyond it, it is in the strong coupling regime, whereas the transverse modes remain weakly coupled above and beyond the strong coupling scale, only with a slight correction due to the longitudinal modes. The modes of the KR field, in contrast, behave opposite. There, the transverse modes become strongly coupled, due to the third term in the Lagrangian density (\ref{KRpert}), at the same strong coupling scale as the self-interacting Proca theory, losing thus its linear propagator. The pseudo-scalar mode on the other hand becomes weakly-coupled even beyond this scale. Thus, even though the two theories describe the same number of dof initially, these modes behave in an entirely different way. The source of this lies in the minimal amplitude of quantum fluctuations for the original transverse and longitudinal modes. 
\\\\
One should note that the two self-interacting theories we have previously considered are not necessarily related with a duality transformation. Nevertheless, if one were to construct a duality transformation, it would be naturally a perturbative one, relating the kinetic terms, and expanding the field in the perturbation theory. However, as we have seen, certain types of self-interactions give rise to the breakdown of the perturbation theory, which means that all transformations that rely on the perturbative expansion around the kinetic terms also fail -- the strongly coupled modes that are absent in the massless theory lose their linear propagator once scales reach the Vainshtein scale. Taking into account also the behavior of the original modes indicates that the duality between the two theories might not hold. 
\section{The non-minimal Proca theory and the tensor mode surprise} \label{nminprocasec}

The previous examples suggest that in principle, dof that appear upon modifying the theory become strongly coupled when the parameter characterizing the modification is very small (tends to zero). Nevertheless, one can still ask -- \textit{How strongly coupled the theory is, and can it allow for the strong coupling of additional modes?} In this section, we will explore this question in Proca theory, with non-minimal coupling to gravity which is described by the following action: 
\begin{equation}\label{PdeBroglieAction}
    S=S_{EH}+S_{P}+S_{nmin},
\end{equation}
where
\begin{equation*}
    \begin{split}
        &S_{EH}=-\frac{M_{pl}^2}{2}\int d^4x\sqrt{-g}R\\
        &S_{P}=\int d^4x \sqrt{-g}\left(-\frac{1}{4}g^{\mu\alpha}g^{\nu\beta}F_{\mu\nu}F_{\alpha\beta}+\frac{m^2}{2}g^{\mu\nu}A_{\mu}A_{\nu}\right)\\
        &S_{nmin}=-\frac{1}{2}\int d^4x \sqrt{-g}\left(\alpha  Rg^{\mu\nu}A_{\mu}A_{\nu}+\beta R^{\mu\nu}A_{\mu}A_{\nu}\right).
    \end{split}
\end{equation*}
This theory has recently attracted a lot of attention in cosmology due to its possibility to realize inflation due to a vector field and generate the primordial magnetic fields \cite{Golovnev:2008cf, Himmetoglu:2008zp, Turner:1987bw, Dimopoulos:2008rf, Dimopoulos:2011ws}. Moreover, it can influence the gravitational particle production of the dark photon \cite{Cembranos:2023qph, Kolb:2020fwh, Ozsoy:2023gnl, Capanelli:2024pzd, Grzadkowski:2024oaf}\footnote{See also \cite{DeFelice:2025ykh} for cosmological implications on the number of dof and their behavior for most general solutions. }. It consists of two types of non-minimal coupling -- one with the Ricci tensor, and one with the Ricci scalar. Yet, even though they are both non-minimal, the two couplings give rise to significantly different effects. In this section, we will study the two in flat space-time separately, by focusing on the type of strong coupling that arises due to them. 
\\\\
In contrast to the previous cases where we have neglected the metric perturbations, the above theory now describes five dof. Three belong to the Proca theory, and two arise due to the linearized gravity\footnote{One should note that an extra dof with vanishing speed of propagation appears if one instead considers an expanding homogeneous and isotropic or anisotropic background and non-vanishing background value of $A_0$, as shown in \cite{DeFelice:2025ykh}.}. Following the same procedure as before, our goal is to study the theory only in terms of the propagating modes. Thus, we will decompose the metric according to
\begin{equation}
    g_{\mu\nu}=\eta_{\mu\nu}+h_{\mu\nu},
\end{equation}
where $\eta_{\mu\nu}$ is the Minkowski metric, and further separate the vector and metric perturbations according to the rotation group: 
\begin{equation}\label{decomposition}
    \begin{split}
        &h_{00}=2\phi\\
        &h_{0i}=B_{,i}+S_i,\qquad\qquad S_{i,i}=0\\
        &h_{ij}=2\psi\delta_{ij}+2E_{,ij}+F_{i,j}+F_{j,i}+h_{ij}^{T},\qquad\qquad F_{i,i}=0,\quad h_{ij,i}^{T}=0,\quad h_{ii}^{T}=0\\
        &A_0\\
        &A_i=A_i^T+\chi_{,i},\qquad\qquad A^T_{i,i}=0.
    \end{split}
\end{equation}
One can easily show that in addition to the temporal component of the vector field, the scalar potentials  $\phi,\psi, B, E$ and vector potentials $S_i$ and $F_i$ are not propagating as well. By finding and solving their corresponding constraints, and substituting them back to the action, we arrive at a theory that contains only the propagating modes, with the linearized Lagrangian given by: 
\begin{equation}\label{linearizednmin}
         \mathcal{L}_{0}=-\frac{1}{2}\left[A_i^T(\Box+m^2)A_i^T+\chi(\Box+m^2)\frac{m^2(-\Delta)}{-\Delta+m^2}\chi\right]+\frac{M_{pl}^2}{8}\partial_{\mu}h_{ij}^T\partial^{\mu}h^{T}_{ij}.
\end{equation}
Following \cite{Hell:2024xbv}, let us now consider the non-minimal couplings separately, focusing on the main results. 
\\\\
As a first step, let us consider the coupling between the vector field and the Ricci scalar. The leading order, which corresponds to the cubic terms in the Lagrangian density, consists only of gravitational scalar modes coupled to the vector field. However, these modes are not propagating. Thus, once of resolve their constraints, they give rise to quartic self-interacting terms for the vector fields. In particular, on the level of the equation of motion, this corresponds to the following expression for the longitudinal modes: 
\begin{equation}
    -m^2(\Box+m^2)\chi\sim\frac{3\alpha^2}{M_{pl}^2}\partial_{\mu}\left[\chi^{,\mu}\Box(\chi_{,\alpha}\chi^{,\alpha})\right]
\end{equation}
Here, we have only presented the most dominant terms. One can notice that this appears to be very similar to the quartic self-interaction of the vector field, even though we have initially started with interactions. By taking into account the minimal amplitude of quantum fluctuations for the longitudinal modes, we find that the leading order corrections will become of the same order as the linear term at the length scales: 
 \begin{equation}
     L_{\alpha}\sim\left(\frac{\alpha}{M_{pl}m^2}\right)^{1/3}.
 \end{equation}
This means that once the strong coupling scale $L_{\alpha}$ is reached, the longitudinal modes become strongly coupled. In contrast, one can verify that tensor and transverse modes remain in the weakly coupled regime up to the Planck scale. 
\\\\
Let us now consider the coupling to the Ricci tensor. In contrast to the Ricci scalar, which at leading order contains only constrained modes, the Ricci tensor includes the tensor ones as well. Thus, its main contributions will include also modes at the cubic order on the level of action. However, by studying the contributions on the level of equations of motion, this coupling introduces a surprising result: the right-hand side of the equation of motion for the tensor modes involves the longitudinal modes, which give rise to singular behavior when perturbatively evaluated: 
\begin{equation}\label{tenold}
    \Box h_{ij}^T\sim -\frac{\beta}{M_{pl}^2}P_{ijkl}^T\Box\left(\chi_{,k}\chi_{,l}\right),
\end{equation} 
where $P_{mnij}^T$ is the is the transverse-traceless projector. By taking into account the amplitude of quantum fluctuations, the leading order contributions can be evaluated as: 
\begin{equation}
    h_{ij}^{(1)}\sim\frac{\beta}{M^2_{pl}m^2L^4}, 
\end{equation}
thus becoming of the same order as the linear term $h_{ij}^{T(0)}\sim\frac{1}{M_{pl}L}$, at length scale:
\begin{equation}
    L_{\beta str}\sim \left(\frac{\beta}{M_{pl}m^2}\right)^{1/3}.
\end{equation} 
It is important to mention that in addition to the tensor modes, one can verify that the longitudinal modes also become strongly coupled at the same scale. Therefore, in contrast to the coupling with the Ricci scalar, which only implied the breakdown of the perturbation theory for the longitudinal modes, the Ricci tensor indicates that in addition, the tensor modes become strongly coupled as well. However, this is unphysical. It implies that regardless of how small the mass of a photon is if the non-minimal coupling with the Ricci tensor is present in the theory, the gravitational waves would quickly lose their linear propagator as a result of strong coupling. One might thus be inclined to simply disregard couplings to the Ricci tensor, and consider the Ricci scalar as the only possible viable coupling in flat spacetime. However, before taking such a conclusion, it is interesting to mention that one faces a similar issue in mYM theory.   
\\\\
In section 2, we have stated that the decomposition of the vector field into the transverse and longitudinal modes is non-linear, given by (\ref{eq::Nonlindec}). However, one could in principle use instead the following decomposition:
\begin{equation}\label{mYMdeclin}
    A_i^a=A_i^{Ta}+\chi_{,i},\qquad\text{where}\qquad A_{i,i}^{Ta}=0.
\end{equation}
Then, as a result, one finds that the transverse modes, which now satisfy: 
\begin{equation}\label{eq::Translineom}
    (\Box+m^2)A_i^{Ta}\sim g\varepsilon^{abc}P_{ij}^T\left[\frac{1}{2}\Box\left(\chi^b\chi^c_{,j}\right)\right],\qquad  P_{ij}^T=\delta_{ij}-\frac{\partial_i\partial_j}{\Delta}
\end{equation}
become strongly coupled, at an energy scale that is smaller than that of the violation of unitarity. Nevertheless, by evaluating the theory at this scale, we can notice that the introduction of the term of the form 
\begin{equation}\label{newcouplingmYM}
    \begin{split}
        A_i^a=A_i^{Ta}+\chi_{,i}^a-g\varepsilon^{abc}(\frac{1}{2}\chi_{,i}^b\chi^c),
    \end{split}
\end{equation}
 removes the smaller-scale strong coupling of the transverse modes and shifts the strong coupling scale. In fact, by studying the new theory with the new coupling, one will again find an unphysical result for the transverse modes, but now appearing at a higher energy scale. The whole procedure will then be repeated, including new couplings to the theory until one reaches the scale that matches the one that we have found in the previous section, and the decomposition becomes modified to (\ref{eq::Nonlindec}). 
\\\\
This connection with mYM theory could also be applied in the case of the Ricci tensor coupling. In particular, as pointed out in \cite{Hell:2024xbv}, it can be inferred by studying the equation of motion for the tensor mode, or even right away at the level of action due to the vanishing of the background value for the curvature in the Minkowski background, that the inclusion of the disformal coupling 
\begin{equation}\label{newdecomposition}
    g_{\mu\nu}=\eta_{\mu\nu}+h_{\mu\nu}-\frac{\beta}{M_{pl}^2}A_{\mu}A_{\nu},
\end{equation}
removes the problematic coupling between the tensor and longitudinal modes, shifting the strong coupling to only the longitudinal modes. It is also interesting to note that, unlike the mYM case, the strong coupling scale that appears for the longitudinal modes after the inclusion of the disformal coupling matches the one that we have found in the original theory. 
\\\\
Recently, it was shown that in addition to the ambiguity in the tensor modes, the Ricci tensor coupling gives an additional issue -- the appearance of runaway modes -- dof in the homogeneous and isotropic universe with negative frequency \cite{Capanelli:2024pzd}. Curiously, however, by including the same form of the disformal coupling, now with 
\begin{equation}
    g_{\mu\nu}=\Tilde{g}_{\mu\nu}-\frac{\beta}{M_{pl}^2}A_{\mu}A_{\nu},\qquad \text{where} \qquad \Tilde{g}_{\mu\nu}=a^2\eta_{\mu\nu},
\end{equation}
such modes disappear. This can be seen from the action of the non-minimal Proca theory, which is in the disformal frame given by: 
\begin{equation}\label{disformalaction}
    \begin{split}
        S=\int d^4x\sqrt{-\Tilde{g}}&\left[\frac{M_{pl}^2}{2}\Tilde{R}-\frac{1}{4}\Tilde{g}^{\mu\alpha}\Tilde{g}^{\nu\beta}F_{\mu\nu}F_{\alpha\beta}-\frac{m^2}{2}\Tilde{g}^{\mu\nu}A_{\mu}A_{\nu}\right.\\&\left.-\frac{1}{2}(\alpha+\beta)\Tilde{R}\Tilde{g}^{\mu\nu}A_{\mu}A_{\nu}+\mathcal{O}\left(\frac{\beta^2}{M_{pl}^2}A^4\Tilde{R}\right)\right]
    \end{split}
\end{equation}
to the leading order. It includes only the Ricci scalar and thus yields only healthy modes if one assumes solutions with vanishing background value of the vector field. 
\\\\
Therefore, by formulating the non-minimal theory in the disformal frame, and ensuring that in this frame the coupling with matter is minimal, thus making the frame physical, one could safely avoid the strong-coupling of the tensor modes as well as the runaway of the longitudinal ones, at least for the cases where the vector field has a vanishing background value. Notably,  the form of the equation that characterizes this disformal coupling suggests that this might even be applied to more general backgrounds. 

\section{Conclusion}

Massive gauge theories with mass introduced by hand to the Lagrangian pose a very intriguing question -- if we allow for such a simple modification, can we smoothly recover the massless theory when this modification is set back to zero? According to the principle of continuity, the answer is a straightforward yes. This is also natural as physics should not be sensitive to if a particle has a mass that is very close to zero or no mass at all. In these notes, we have discussed examples where this question was, however, not trivial to answer. 
\\\\
By studying the massive Yang-Mills theory, with mass added by hand, we have shown that perturbatively, it would seem that the series is singular in mass, due to the appearance of the longitudinal mode -- a degree of freedom that is absent in the massless theory. However, by studying the equations of motion, we have found that due to the non-linear corrections, the same mode becomes strongly coupled. As a result, it loses its linear propagator and decouples from the remaining transverse modes up to very small corrections that ultimately vanish in the massless limit. The transverse modes, on the other hand, remain always in the weakly coupled regime. This result shows a conceptual connection with massive gravity, which also suffers from an apparent discontinuity due to the longitudinal modes, which is resolved beyond the perturbation theory. 
\\\\
Notably, both of these theories suggest a trend -- dof that appears upon modifying a theory will become strongly coupled when the modification is taken back to zero. By exploring two other theories, of self-interacting Kalb-Ramond and Proca fields, we have found further support for this claim. In particular, Proca behaves in a similar way as its non-Abelian version -- the longitudinal mode which is absent in the massless theory becomes strongly coupled due to the presence of self-interaction. Beyond this scale, it then decouples from the remaining degrees of freedom, the transverse modes, which in turn remain in the weakly coupled regime.  In the Kalb-Ramond case, on the other hand, the transverse modes become strongly coupled, while the pseudo-scalar remains weakly coupled. Thus, while the two theories describe the same number of dof, the same yields different behavior in the two theories, raising a question on their previously thought connections. 
\\\\
By exploring the Proca theory with non-minimal coupling to gravity we have arrived at another important aspect -- how strongly coupled the theory can be. While in the previous three examples, only modes that are absent in the massless case are strongly coupled, while the remaining ones remain weakly coupled for all scales, the latter example brings a surprise -- in addition to the longitudinal mode, the perturbation theory also breaks down for the tensor modes, thus resulting in an unnatural behavior. Nevertheless, by introducing disformal couplings to the theory, in a similar way as in the mYM theory, we have shown that the problematic coupling can be removed and render the theory consistent. 
\\\\
Overall, while such theories are not straightforward to analyze as perturbative tools are not applicable for all of the modes beyond the strong coupling scale, they nevertheless provide interesting and curious models, especially in cosmological and black-hole applications, which may lead to further surprises and interesting phenomena. 

\begin{center}
    \textbf{Ackgnowledgments}
\end{center}
\textit{ I would like to thank the organizers of the CORFU2024, and The Dark Side of the Universe - DSU2024, for inviting me to give a plenary talk based on which these notes were given. In particular, I would also like to thank to George Zoupanos for organizing the wonderful CORFU workshops for many years. Also, I would like to thank all of the physicists with whom I had great discussions about these topics, and in particular to my PhD advisor Slava Mukhanov, and Dieter L\"ust, Elisa Ferreira, Misao Sasaki, Arkady Vainshtein and George Zoupanos, for support and essential discussions while this research works were done. The work of A. H. was supported by the World Premier International Research Center Initiative (WPI), MEXT, Japan. 
 }

\end{document}